\documentclass[a4paper,aps,superscriptaddress,floatfix,nofootinbib,twocolumn,bibtex]{revtex4}

\usepackage{bbold}
\usepackage{bbm}
\usepackage[pdftex]{graphicx}
\usepackage{latexsym,amsmath,verbatim}
\usepackage{color}
\usepackage{rotating}
\usepackage{verbatim}
\usepackage{multirow}
\usepackage[english]{babel}
\usepackage{comment}

\begin{document}

\title{Breaking of the site-bond percolation universality in networks}

\author{Filippo Radicchi}
\affiliation{Center for Complex Networks and Systems Research, School of Informatics and Computing, Indiana University, Bloomington, USA}
\email{filiradi@indiana.edu.}

\author{Claudio Castellano}
\affiliation{Istituto dei Sistemi Complessi (ISC-CNR), Roma, Italy, and
Dipartimento di Fisica, Sapienza Universit\`a di Roma, Roma, Italy}

\begin{abstract}
  The stochastic addition of either
  vertices or connections in a network leads to
  the observation of the percolation transition,
  a structural change with the appearance of a
  connected component encompassing a finite fraction
  of the system. Percolation has always been regarded as
  a substrate-dependent but model-independent
  process, in the sense that the critical exponents of the
  transition are determined by the geometry of the system, 
  but they are identical for the bond and site percolation models.
  Here, we report a violation of such assumption.
  We provide analytical and numerical
  evidence of a difference in the values of the critical
  exponents between the bond and site percolation models
  in networks with null percolation thresholds, such as 
  scale-free graphs with diverging
  second moment of the degree distribution.
  We discuss possible implications of our
  results in real networks, and provide additional insights
  on the anomalous nature of the percolation transition
  with null threshold.
\end{abstract}

\maketitle

\section{Introduction}

Percolation is among the simplest processes
able to generate continuous phase
transitions~\cite{stauffer1991introduction, bollobas2006percolation}.
The model used to describe percolation assumes the presence of an underlying
network structure where either nodes (site percolation) or edges (bond
percolation) are randomly occupied with probability $p$. Nearest-neighbor
occupied elements form connected clusters.
In site
percolation, for $p=0$, no elements are
present in the system, so that all clusters
have size zero. In bond percolation, for $p=0$,
no nodes are connected in the system,
so that all clusters have size equal to one. In both
models, for $p=1$, only a single
cluster, coinciding with the whole network, 
is present. The term 
percolation transition refers to the structural 
change, between these two extreme configurations, 
observed as a function
of the occupation probability $p$.
The change is usually monitored through the relative
size of the largest cluster, or percolation strength, which is
regarded as the order parameter
of the percolation transition. 
In the limit of infinitely large networks, 
this observable is always equal to zero
for any value of $p \leq p_\textrm{c}$, while it is finite for $p > p_\textrm{c}$. 
Whereas the percolation threshold $p_\textrm{c}$ can be different in the
two models, for a fixed underlying
network, bond and site percolation processes have been
always observed to behave identically around their
respective threshold values. The 
exponent describing the power-law growth
of the order parameter as a function of the
distance from the critical point is
the same in both processes~\cite{stauffer1991introduction}.
This statement is true also for the critical exponents 
that describe the singular behavior of other observables, such as the
distribution of the cluster size, and the average size 
of finite clusters.
The specific values of the 
critical exponents play an important role in the
characterization of the properties of the
percolation transition,
and they are used to group networks in
different universality classes. 
In lattices for example,
the values of the critical exponents depend only 
on the dimensionality of the euclidean space~\cite{stauffer1991introduction}.
Such a dependence disappears above the upper-critical
dimension, where the
critical exponents stabilize to their mean-field 
values~\cite{stauffer1991introduction}.
In random networks also, no differences have been reported 
between the critical exponents of the bond and site percolation
models~\cite{callaway2000network, moore2000exact,
cohen2002percolation,dorogovtsev2008critical}. 
Theoretical approaches proposed so far indeed assume
a perfect equivalence between the models~\cite{dorogovtsev2008critical}.
In this paper we are going to show that this assumption is incorrect.
In graphs with null percolation threshold, as for example
random networks with diverging second moment
of the degree distribution, bond and site
percolation strengths are characterized by different
critical exponents. The breaking of the site-bond universality is 
accompanied with
anomalies in the critical behavior of other macroscopic observables.

\section{Results}

\subsection{Bond percolation model}
We first derive the basic equations that support our statement, 
starting from the bond percolation model.
We assume the presence of an underlying undirected
and unweighted network composed of $N$ nodes
and $E$ edges. The structure of the network
is fully described by the adjacency matrix $A$.
The generic  element of this matrix equals
one if the two corresponding nodes share an edge, whereas
equals zero if no connection is present between the
two vertices. The probability $b_i$ that node $i$ is
part of the largest cluster of the network is a
function of $A$ and the bond occupation probability $p$.
Such a probability obeys the equation
\begin{equation}
  b_i = 1 - \prod_{j \in \mathcal{N}_i} \, ( 1 - p \, c_{i \to j} ) \;.
  \label{eq:bond_sparse1}
\end{equation}
 Here, $\mathcal{N}_i$ is the set of neighbors
of vertex $i$, while $c_{i \to j}$ stands for the 
probability that node $j$
is part of the largest cluster discounting 
the contribution of node $i$.
Eq.~(\ref{eq:bond_sparse1}) 
is formulated according
to the following straightforward argument.
If node $j$ is in the set $\mathcal{N}_i$ of neighbors
of vertex $i$, then $p \, c_{i \to j}$ is the probability
that the connection between $i$ and $j$
is occupied, and node $j$ is part of the
spanning cluster thanks to a
node different from $i$. Thus, the probability
that node $i$ does not belong to the largest cluster,
i.e., $1 - b_i$, is equal to the probability
that none of its adjacent nodes, that
are connected to vertex $i$ by an occupied edge, are
part of the largest cluster of the graph.
Note that Eq.~(\ref{eq:bond_sparse1}) is based on the
hypothesis that the probabilities
$c_{i \to j}$ of all neighbors
of node $i$ are uncoupled, i.e., the so-called
locally tree-like approximation~\cite{dorogovtsev2008critical}, hence
their product appears on the r.h.s. of
the equation.
For consistency, the
probability $c_{i \to j}$ obeys
\begin{equation}
  c_{i \to j} = 1 - \prod_{k \in \mathcal{N}_j \setminus \{i\}} \, ( 1 - p \, c_{j \to k} )\;, 
  \label{eq:bond_sparse2}
\end{equation}
where the product on the r.h.s. of the last equation runs
over all neighbors of node $j$ but vertex $i$.
Given the adjacency matrix $A$ of the
underlying graph, and fixed a value
of the occupation probability $p$,
the solution of the bond percolation model
can be obtained first by numerically solving the set
of $2E$ Eqs.~(\ref{eq:bond_sparse2}), and then plugging these
solutions into the set of $N$ Eqs.~(\ref{eq:bond_sparse1})
to estimate the value of the variables $b_i$.
The order parameter
of the transition can be finally computed as 
the average value of these variables over
the entire network, i.e., $B = 1/N \, \sum_i b_i$.
This quantity represents the percolation strength
$B$ over an infinite number of realizations of the
bond percolation model on the graph.
Using the Taylor expansion of Eqs.~(\ref{eq:bond_sparse2})
around $c_{i \to j} = 0$, it
can be shown that the percolation threshold
equals the inverse of the largest eigenvalue of the
non-backtracking matrix of the graph~\cite{PhysRevLett.113.208702},
and that
slightly on the right of the critical probability, every
$b_i$ grows linearly with the sum of the components of the
principal eigenvector of the
non-backtracking matrix corresponding to edges pointing out
from node $i$~\cite{radicchi2015percolation}.

\subsection{Site percolation model}
Under the locally tree-like approximation, the
probability $s_i$ that node $i$ belongs to the
largest cluster in the network is given by
\begin{equation}
  s_i = p \,[\, 1 -  \prod_{j \in \mathcal{N}_i} \, ( 1 - t_{i \to j}) \, ] \; ,  
  \label{eq:site_sparse1}
\end{equation}
where $t_{i \to j}$ stands for the probability that node $j$
is part of the largest cluster irrespective
of vertex $i$. The probability $s_i$
is written as the product of two contributions:
(i) the probability $p$ that the node
is occupied; (ii) the probability that at least one
of its neighbors is part of the largest cluster independently
of node $i$. For consistency, the
probability $t_{i \to j}$ obeys
\begin{equation}
  t_{i \to j} = p \, [ 1 - \prod_{k \in \mathcal{N}_j \setminus \{i\}} \, ( 1 -  t_{j \to k}) \, ] \; ,
  \label{eq:site_sparse2}
\end{equation}
where we have excluded node $i$ from the product on the r.h.s.
As in the case of bond percolation, Eqs.~(\ref{eq:site_sparse2})
form a set of $2E$ coupled equations whose solution can be
obtained numerically for any value of $p$. The numerical solutions
of Eqs.~(\ref{eq:site_sparse2}) are then plugged into 
Eqs.~(\ref{eq:site_sparse1}) to obtain the values of the variables $s$, and
finally the order parameter of the transition is computed 
as $S = 1/N \, \sum_i s_i$. Also in this case, the percolation threshold
equals the inverse of the largest eigenvalue of the
non-backtracking matrix of the graph~\cite{PhysRevLett.113.208701}.

%%%%%%%%%%%%%%%%%%%%%%%%%%%
\begin{figure}[!htb]
  \begin{center}
    \includegraphics[width=0.5\textwidth]{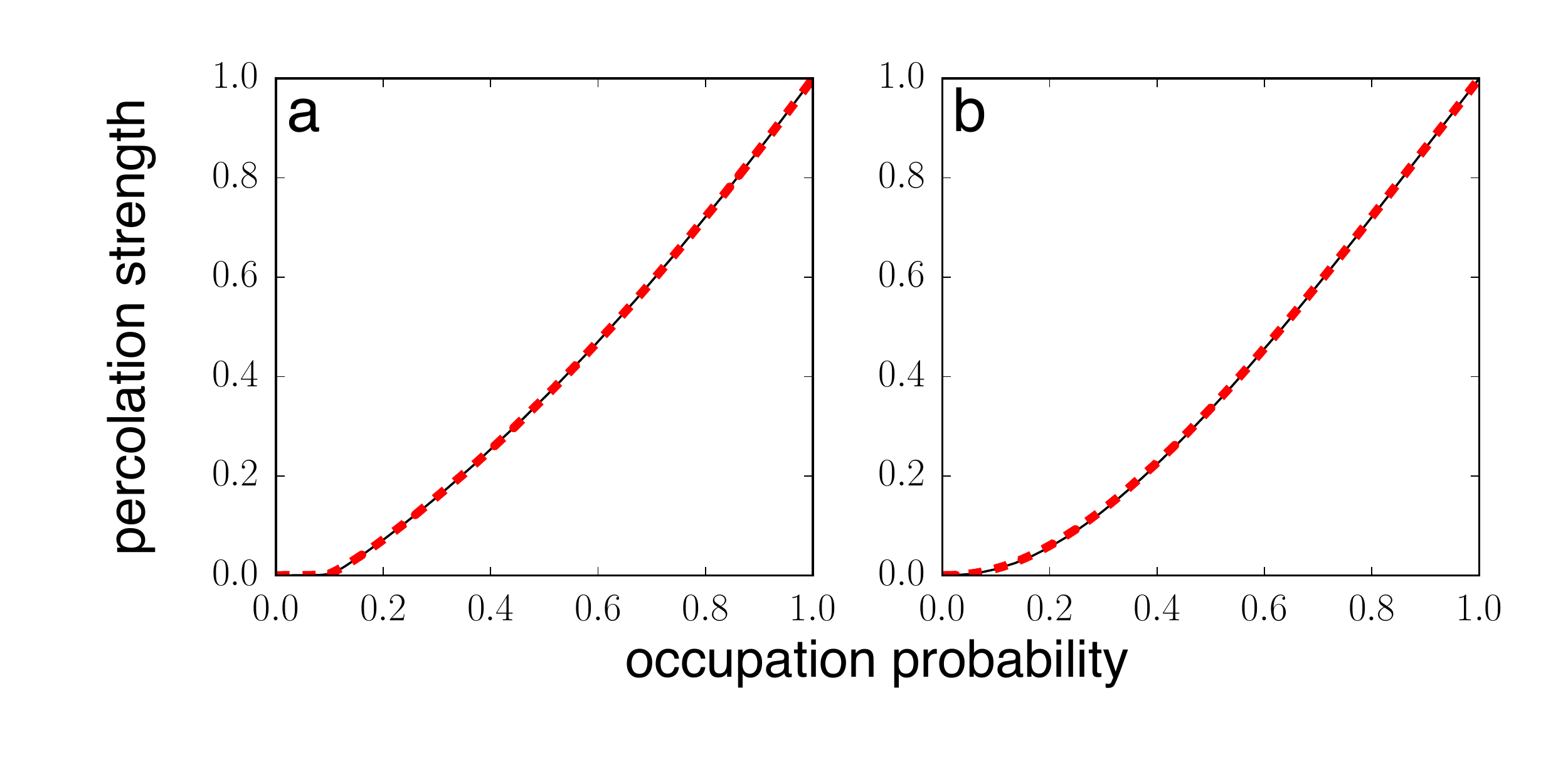}
  \end{center}
    \caption{
      Percolation diagrams of real networks.
      ({\bf a}) We consider the graph corresponding to the giant
      component of the {\it peer-to-peer Gnutella}
      network as of August 31, 2002~\cite{ripeanu2002mapping,
        leskovec2007graph}.
      The black thin line represents
      the site percolation order parameter $S$ as a function
      of the site occupation probability $p$.
      We calculate also the order parameter $B$ for
      bond percolation and multiply
      it by $p$ to obtain the red dashed line.
      The
        average clustering coefficient of the network is $C = 0.0055$.
        Such a low value
        indicates that the tree-like approximation
        holds with sufficient accuracy for this network.
        We further estimated the error $V$ of the law $S = p B$,
        by considering the integral $V = \int_{0}^1 \, dp \; |S(p) - p B(p)|$.
        We find $V = 0.0002$.
      ({\bf b}) We consider the graph
      corresponding to the giant
      component of the {\it Internet at the autonomous system
        level} in the period
      January 2004 to November
      2007~\cite{leskovec2005graphs}. The description
      of the various curves is identical
      to those appearing in panel a. The clustering
      coefficient for this network is 
      $C = 0.2082$. The error associated to Eq.~(\ref{eq:main}) is $V = 0.0016$.
    }
\end{figure}
%%%%%%%%%%%%%%%%%%%

\subsection{Relation between bond and site percolation} If we multiply both sides
of Eq.~(\ref{eq:bond_sparse2}) by $p$, we recover
Eq.~(\ref{eq:site_sparse2}), with
only the necessity of renaming 
$p  c_{i \to j} \to t_{i \to j}$. The same is also true
for Eq.~(\ref{eq:bond_sparse1}) which reduces to Eq.~(\ref{eq:site_sparse1})
with a multiplication by $p$, and the additional change of variable 
$p \, b_i \to s_i$. As a consequence, the
percolation strengths $B$ and $S$ are related by
\begin{equation}
  S = p \, B \; ,
  \label{eq:main}
\end{equation}
which tells us that, in locally tree-like networks,
  the order parameters of the bond and site percolation
  models are linearly proportional~\cite{PhysRevE.91.052807}. 
Eq.~(\ref{eq:main})
holds with very high accuracy in 
many real networks,
as long as their structure is 
sufficiently compatible with
the locally tree-like
approximation [see Fig.~1, 
  Supplementary Figs.~1-109, Supplementary
  Tables~1-3, and Supplementary Note 1 for results on $109$ 
real networks~\cite{PhysRevE.91.010801}]. To provide
  a quantitative test of this statement,
  we estimate 
  the error associated to Eq.~(\ref{eq:main}) as
  $V = \int_{0}^1 \, dp \; |S(p) - p B(p)|$, and
  use the average
  clustering coefficient $C$ as a proxy for the validity of
  the tree-like ansatz. 
  We find $V<0.1$ for all the real networks we analyzed (see
  Supplementary Tables~1-3),
  suggesting a good accuracy of Eq.~(\ref{eq:main})
  overall. For most networks with relatively
  low values of the clustering coefficient,
  Eq.~(\ref{eq:main})
  works exceptionally well (i.e., $V<0.01$). 
  On the other hand, we find also a
  positive dependence of $V$ on $C$, indicating
  that the accuracy of Eq.~(\ref{eq:main}) decreases
  as the tree-like approximation becomes less reliable
  (see Supplementary Fig.~110).

\subsection{Violation of the site-bond percolation universality} 
From Eq.~(\ref{eq:main}) a difference 
in the critical behavior between the bond and site percolation
models is straightforwardly deduced. 
In infinitely large networks, as the occupation
probability tends to the critical
threshold value from right, i.e., $p \to p_\textrm{c}^+$, the
order parameter of the percolation transition
decreases to zero as a power of the distance from the
critical point, that is
$B \sim (p - p_\textrm{c})^{\beta_\textrm{b}}$ and $S \sim (p - p_\textrm{c})^{\beta_\textrm{s}}$.
Whereas in the former equations we stressed
the possibility of a difference in the values
of the critical exponents for the bond and site
percolation models, we remark that there
are not known examples of such observation.
On the contrary, it is firmly believed that
the value of critical exponents
depends only on the geometry of the system
but not on the specific ordinary
percolation model considered~\cite{stauffer1991introduction}.
By making use of the linear mapping
of Eq.~(\ref{eq:main}), we can write
$p (p -p_\textrm{c})^{\beta_\textrm{b}} \sim (p - p_\textrm{c})^{\beta_\textrm{s}}$.
If the percolation threshold is strictly larger
than zero, as in the case of regular graphs,
Erd\H{o}s-R\'enyi models, or random scale-free graphs
with finite second moment of the degree
distribution, in the limit $p \to p_\textrm{c}^+$, the prefactor
$p$ on the l.h.s. of the previous
equation acts as a multiplicative constant, and
$\beta_\textrm{b} = \beta_\textrm{s}$. 
If instead $p_\textrm{c} = 0$, as in the case of
random scale-free graphs with diverging second moment
of the degree distribution~\cite{albert2000error,
  callaway2000network, cohen2000resilience},
the former equation
becomes $p^{\beta_\textrm{b} + 1} \sim p^{\beta_\textrm{s}}$.
The critical exponents of the percolation
strengths of bond and site
percolation are thus related by
\begin{equation}
  \beta_\textrm{s} = \beta_\textrm{b} + 1 \; , 
  \label{eq:violation}
\end{equation}
which tells us that, in locally tree-like graphs
  with null percolation thresholds, the site-bond 
  universality is broken,
  and the critical exponents of the order parameters of the
  bond and site percolation models assume different values.

  %%%%%%%%%%%%%%%%%%%%%
\begin{figure}[!htb]
  \begin{center}
    \includegraphics[width=0.5\textwidth]{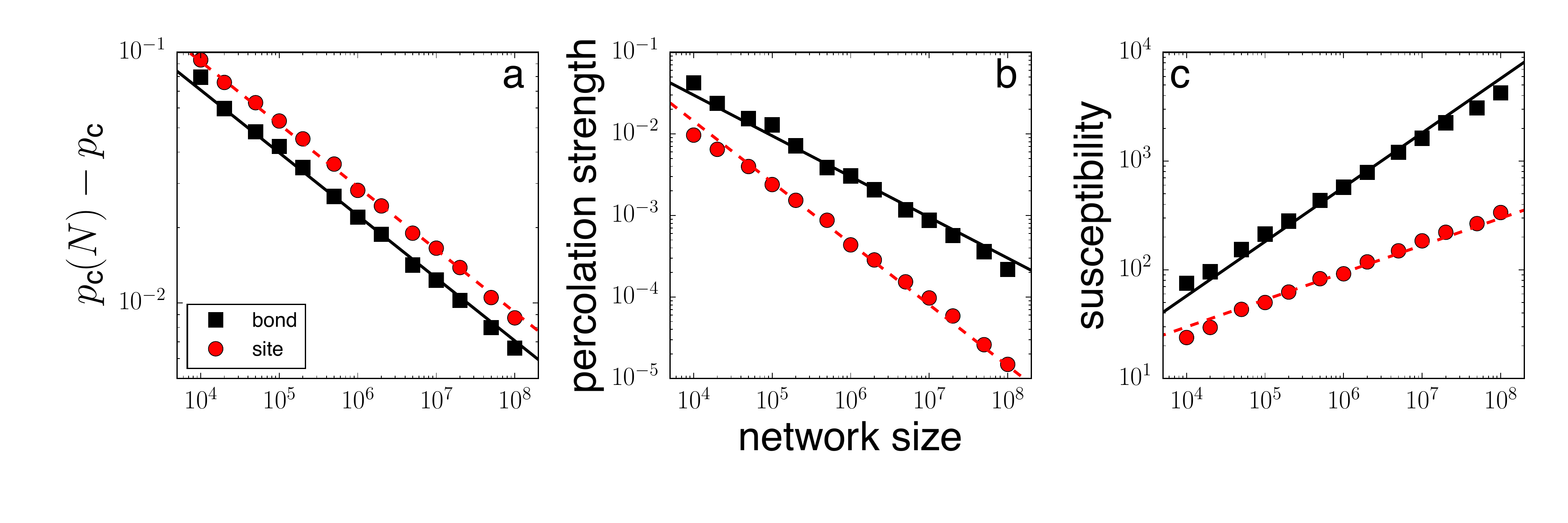}
  \end{center}
  \caption{
      Critical behavior of bond  and site 
      percolation on
      scale-free graphs.
    Results are obtained for random networks with degree 
    distribution $P(k) \sim k^{-\gamma}$ built according to the uncorrelated
    configuration model (see methods) and setting the degree exponent 
    $\gamma = 2.5$. Black squares 
    refer to bond percolation, while red circles
    represent the results obtained for site percolation.
    ({\bf a}) Best estimate of the
    pseudo-critical point $p_\textrm{c}(N)$ 
    for different network sizes $N$. 
    Simulation results (symbols) are compared with 
    the expected power-law decay (full black and dashed red lines
    are guides to the eye)
    towards $p_\textrm{c}$, with
    $p_\textrm{c} = 0 $ and decay exponent $1/\nu = (3 - \gamma)/2$.
    ({\bf b}) Percolation strengths $B$  (black squares)
    and $S$ (red circles) at $p=p_\textrm{c}(N)$
    as functions of the network size $N$. The full black line serves as a
    guide to the eye and decays 
    with exponent $\beta_\textrm{b}/\nu = 1/2$ as $N$ grows (see methods).
    The dashed red line serves as a
    guide to the eye to indicate a power-law decay with an exponent
    $\beta_\textrm{s}/\nu=(4-\gamma)/2$ (see methods).
    ({\bf c}) Maximal values of susceptibilities $\chi_\textrm{B}$ (black squares)
    and $\chi_\textrm{S}$ (red circles) as functions of $N$. 
    The full black line increases as $N^{1 - \beta_\textrm{b}/\nu}$, thus as $N^{1/2}$.
    The dashed red line stands for guide to the eye
    for the power-law divergence
    $N^{1 - \beta_\textrm{s}/\nu}$, which means $N^{(\gamma-2)/2}$.
  }
\end{figure}
%%%%%%%%%%%%%%%%%%%

\begin{figure}[!htb]
  \begin{center}
    \includegraphics[width=0.5\textwidth]{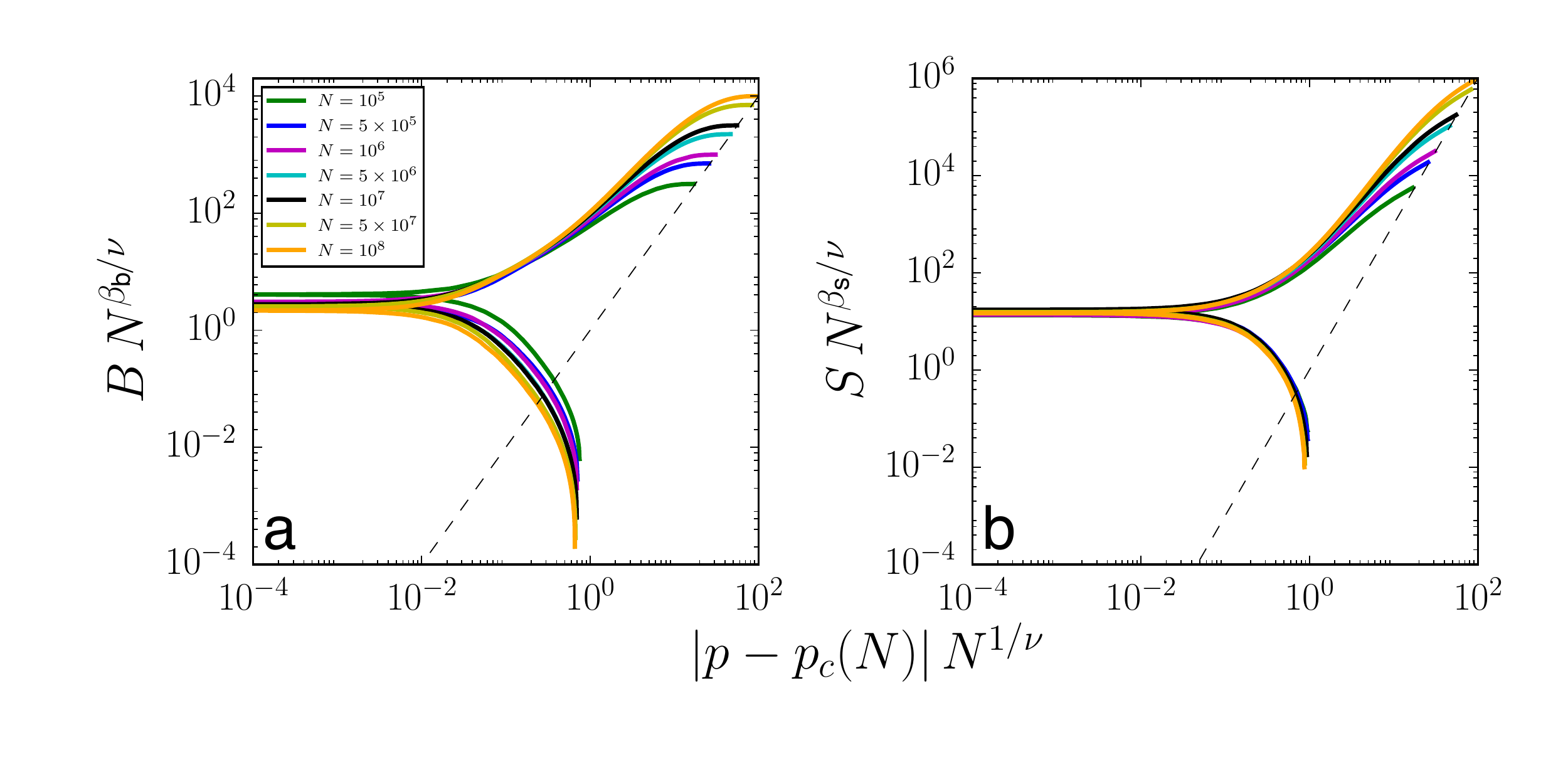}
  \end{center}
  \caption{
    Finite-size scaling in scale-free graphs.
    We analyze the same networks as in Fig.~2, and 
    test the validity of Eq.~(\ref{eq:finite}).
    ({\bf a}) Collapse plot for the order parameter
    $B$ of the bond percolation model in networks
    with different sizes. The collapse is obtained
    by setting $\beta_\textrm{b}/\nu = 1/2$ and
    $\nu = 2/(3-\gamma)$. The dashed line
    corresponds to a guide to the eye for a power-law with
    exponent equal to $\beta_\textrm{b}= 1/(3 - \gamma)$.
    The imperfect collapse for $|p-p_\textrm{c}(N)| N^{1/\nu} \ll 1$ is
    a consequence of the effective exponent $\beta_\textrm{b}/\nu$ in Fig.~2b,
    which is slightly larger than the value $1/2$ due to preasymptotic
    effects.
    ({\bf b}) Same as in panel
    a, but for the site percolation model. In this case,
    we set $\beta_\textrm{s}/\nu=(4-\gamma)/2$ and
    $\nu = 2/(3-\gamma)$. The dashed line is a guide to the eye for
    a power-law with exponent equal
    to $\beta_\textrm{s} = (4-\gamma)/(3-\gamma)$.
  }
\end{figure}
%%%%%%%%%%%%%%%

To validate our theoretical predictions,
we numerically study the two percolation models
in random graphs~\cite{molloy1995critical, catanzaro2005generation}
using the Monte Carlo
algorithm introduced by Newman and
Ziff~\cite{newman2000efficient}.
We consider random network models
that are sparse enough to satisfy
the locally tree-like ansatz, and extrapolate critical
exponent values of the transition
for networks of infinite size
by making use of finite-size scaling
analysis~\cite{stauffer1991introduction}.
First, we verify that for random graphs
with nonvanishing percolation thresholds
identical values for the critical exponents in bond
and site percolation are indeed recovered (Supplementary Figs.~111-118).
In particular, for networks
with power-law degree distribution but finite second moment,
we obtain values of the critical exponents
consistent with previous theoretical
predictions~\cite{cohen2002percolation, wu2007numerical}.
These statements are
valid not just for the critical exponent $\beta$, but also
for the one that regulates the divergence of the
average cluster size, as well as for the Fisher exponent
of the distribution of cluster sizes at
criticality~\cite{stauffer1991introduction}.
Results for scale-free graphs with
diverging second moment of the degree distribution, and thus
null percolation thresholds, are reported in Fig.~2. 
For the bond percolation model, we recover the value of the exponent
$\beta_\textrm{b}$ predicted by the theory of Cohen {\it et al.}
~\cite{cohen2002percolation}.
For the site percolation model, we find instead results consistent 
with our Eq.~(\ref{eq:violation}) (see also 
Refs.~\cite{dorogovtsev2008critical, dorogovtsev2010lectures}).
These different values of the critical
exponents are confirmed in Fig.~3 by the good scaling collapse
among curves corresponding to different network sizes.

\subsection{Interpretation of universality breaking} 
What is the physical reason of the
  difference between the exponents $\beta$ in the
  two percolation models? To get
  insights, consider
  a star-like graph, where a single node is connected to
  an infinitely large
  number of vertices. This structure represents the
  extreme limit of a network with
  diverging second moment of the degree
  distribution, and it is often used to understand
  basic mechanisms induced by the heterogeneity of the
  node degrees~\cite{dorogovtsev2008critical}.
  In the bond percolation
  model, every node at the end of an active edge
  is automatically part of the largest cluster.
  An increment in the occupation probability $p$
  generates a linear increment of the relative size of the
  largest cluster, that is  $B \sim p$. In the site percolation 
  model instead, the largest cluster can grow only if the center
  of the star is active. This happens with probability $p$. If the center
  of the star is active, then the growth of the largest cluster
  is determined by the total number of other vertices that
  are active, that is the rate of growth of
  the largest cluster in the bond percolation model.
  Thus, the relative size of the largest
  cluster in the site percolation model
  behaves as $S \sim p^2$, in accordance
  with Eq.~(\ref{eq:violation}).
  
  We expect the same physical principle
  to play a fundamental role
  in percolation processes on random graphs
  with degree distribution $P(k)$.
  The giant connected component, near its point of creation,
  has degree distribution  proportional to $k P(k)$, hence it consists
  mostly of vertices with high degrees or
  hubs~\cite{dorogovtsev2008critical}
  (Supplementary Fig.~119).
  Bond and site percolation models differ, however,
  in the way nodes with different
  degree become part of connected clusters.
  In the bond percolation model,
  there is a preference for selecting edges
  attached to hubs, as
  $k \, P(k)$ is the probability
  that a node at the end of a randomly
  selected edge has degree
  equal to $k$. In the site percolation model instead, a node
  with degree $k$ is activated with probability $P(k)$ so
  that there is a weaker preference to select hubs. However, 
  when a high-degree vertex is activated,
  many edges are activated simultaneously, and many
  clusters can be merged together. Such a microscopic difference
  among the two models becomes apparent, with
  different values of the critical exponent $\beta$,
  only if the number of hubs
  is sufficiently large, as for example
  in scale-free graphs with $P(k) \sim k^{-\gamma}$
  and degree exponent $2< \gamma <3$.
  For $\gamma > 3$ instead, 
  hubs are too rare to generate differences at the
  macroscopic level, and the site-bond percolation
  universality is restored.

  %%%%%%%%%%%%%%%%%%%%%%%
\begin{figure}[!htb]
  \begin{center}
    \includegraphics[width=0.5\textwidth]{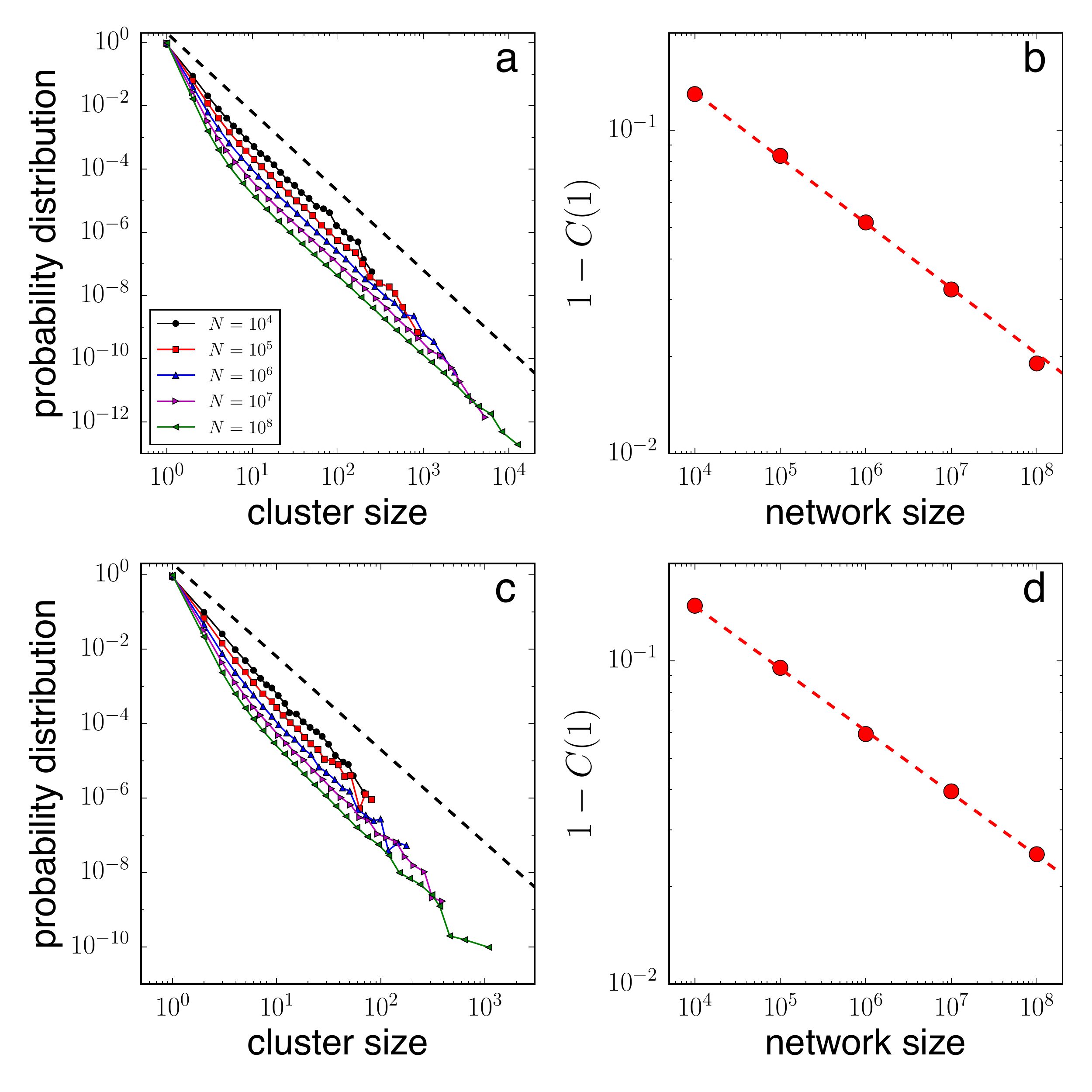}
  \end{center}
  \caption{
    Distribution of finite-cluster sizes
      in scale-free graphs.
    We analyze the same networks as in Figs.~2 and 3.
    ({\bf a}) Probability distribution to observe
    a cluster of a given size in the bond
    percolation model for $p = p_\textrm{c}(N)$.
    Each curve
    corresponds to a different
    network size. The tail of the various
    distributions decays as a power-law
    for large values of the cluster sizes
    with exponent compatible with $5/2$ (the dashed 
    line is a guide to the eye)
    ({\bf b}) Weight of clusters of
    size one, $C(1)$, in the distribution of
    finite-cluster sizes. As the system size grows,
    $C(1)$ tends to one in a power-law fashion
    (the red dashed line represents the best power-law fit of
    the empirical points and has decay exponent equal to $0.20$).
    ({\bf c} and ({\bf d}) Same as in panels a and b, but for the
    site percolation model. In panel c, 
    the black dashed line serves as a guide to the
    eye for a power-law decay with exponent $5/2$. In panel d,  the red dashed line
    represents 
    the best power-law fit of
    the empirical points and has decay exponent equal to $0.18$.
  }
\end{figure}
%%%%%%%%%%%%%%%%%%%%%%%%

\subsection{Anomalies of percolation in scale-free graphs} 
The study of other macroscopic observables reveals
that random networks with null percolation
thresholds show anomalies not
just at the level of the critical exponents
of the order parameter, but
in the nature of the transition itself.
In ``standard'' percolation transitions,
the distribution of finite cluster sizes
decreases at criticality as a power-law with
an exponential cut-off diverging as the system
becomes infinite~\cite{stauffer1991introduction}.
In scale-free graphs with null percolation threshold the power-law 
decay is only a preasymptotic effect, visible
only in finite-size systems. This is clearly seen in
Figs.~4a and c, showing a power-law tail which 
tends to disappear in the limit of infinitely large networks. 
The vanishing of the 
power-law tail is confirmed in Figs.~4b and d, showing that all the
distribution weight gets concentrated on clusters of size 1.
This finding is in stark contrast with all theoretical predictions 
proposed so far~\cite{albert2002statistical, cohen2002percolation,
dorogovtsev2008critical, PhysRevE.76.045101},
which are inconsistent with each other,
as they all provide different estimates for the
Fisher critical exponent. We emphasize
that their validity has been never systematically tested in 
numerical experiments. 
Our results can be interpreted by intuitive arguments.
If the percolation threshold is zero, then
the critical configuration is given by a disconnected
network where all clusters have size one
  in bond percolation, and size zero in site percolation.
Analogous considerations 
  about the critical configuration have been deduced for
self-similar graphs~\cite{serrano2011percolation}, although
  no difference between site and bond percolation
  was studied.
As a matter of fact, the Fisher critical exponent is not 
clearly defined, 
because the entire cluster size distribution does not decay as 
a power-law. The same argument implies also that
  the average size of finite clusters
   does not diverge at criticality and its associated critical exponent
  is equal to zero (Supplementary Figs.~120 and~121).

\section{Discussion}

The breakdown of site-bond percolation universality
in locally tree-like networks with null
thresholds is a surprising result. Although
percolation processes have been extensively studied in 
the last decades, to the best of our knowledge,
there are no previous findings of such
discrepancy between the bond and site percolation models.
A relation analogous to Eq.~(\ref{eq:violation}) has been found long ago 
in continuum percolation models for conductivity in $d$-dimensional 
porous rocks~\cite{Trugman86}. We stress however that the similarity 
is only formal, as here the relation is between
standard bond and site percolation, while Ref.~\cite{Trugman86}
connects the $\beta$ exponents of an ordinary and a suitably modified
continuum percolation process in $d$-dimensional spaces.
Our results could have therefore great importance in
percolation theory by stimulating
further research in a direction not yet
explored. Also, we remark that scale-free
graphs with diverging second moments of the degree
distribution are regarded as prototypical
models of a large variety of
natural and man-made networks~\cite{barabasi1999emergence,
  albert2002statistical}.
In this context, our results could have direct
consequences in all situations where percolation plays a
fundamental role, including
spreading processes in
networks~\cite{pastor2001epidemic, newman2002spread,
  pastor2014epidemic},
as well as resilience
properties of graphs to random
breakdowns~\cite{albert2000error, cohen2000resilience,
  kitano2004biological}.
One may remark that many real networks
  are characterized by high values
  of the clustering coefficient~\cite{watts1998collective}, and thus
  violate the tree-like approximation at the basis
  of our mathematical framework. We argue that
  a nonvanishing clustering coefficient is not a sufficient ingredient
  to restore the percolation universality
  class in networks with diverging second moment
  of the degree distribution. By repeating our numerical
  experiments on the generalization of the
  configuration model proposed by Newman~\cite{newman2009random},
  that creates random
  scale-free networks with nonvanishing clustering
  coefficients, we find in fact
  that the anomalous phenomenology
  still persists (Supplementary Fig.~122).
  Other ingredients seem thus necessary to
  observe a nonvanishing percolation threshold
  and consequently to restore the percolation
  universality class in networks
  with diverging second moment of the degree
  distribution. For instance, we
  expect that scale-free network models characterized
  by spatial embedding~\cite{PhysRevE.66.056105} or high
  density of cliques~\cite{PhysRevE.81.066114} will not exhibit
  such an anomalous behavior.

\begin{acknowledgments}
We thank Marian Bogu\~{n}\'{a}, Alessandro Flammini, 
and Romualdo Pastor-Satorras 
for a critical reading of the manuscript.
\end{acknowledgments}

\section{Methods}

\subsection{Order parameters and critical exponents}
The main order parameter used in the study of the
percolation transition in networks is the so-called
percolation strength, defined as the
number of nodes
belonging to the largest connected cluster of the
network divided by the total number of vertices in the graph.
In our paper, we indicated this quantity as
$B$ for bond percolation, and $S$ for site
percolation.
In the limit of infinitely large systems, the 
order parameter $B$ grows as a power-law
function of the distance between the actual value
of the occupation probability $p$ and the
critical threshold $p_\textrm{c}$, that is
\begin{equation}
B \sim (p - p_\textrm{c})^{\beta_\textrm{b}} \; .
\end{equation}
The same behavior is valid for $S$, and the
critical exponent is denoted as $\beta_\textrm{s}$.
As already explained
in the text, the value of the critical
exponents $\beta_\textrm{b}$ and $\beta_\textrm{s}$
is the same 
if the percolation threshold $p_\textrm{c}$ is strictly
larger than zero.
Whereas $B$ and $S$ are based only on the size
of the largest connected cluster in the network, there are
other important macroscopic observables that
account for the size of the other clusters, and
critical exponents that are associated with them.
In our paper, we considered the distribution
of the cluster size at criticality which leads
to the definition of the Fisher exponent $\tau$, and the
average size of finite clusters with associated
critical exponent $\omega$.

\subsection{Numerical simulations}
Given an undirected and unweighted network with
$N$ nodes and $E$ edges composed of a single connected
component, we study bond percolation using the Monte
Carlo method proposed by Newman and Ziff~\cite{newman2000efficient}.
In each
realization of the method, we start from a configuration with no
connections. We then sequentially add edges in random order
and monitor the evolution of the size of the largest cluster
in the network $Z(p)$ as a function of the bond occupation
probability $p = e/E$, where $e$ indicates the number of edges
added from the initial configuration, i.e., $e = 0$.
We repeat the entire process $Q$ independent times and estimate
the order parameter $B$ as
\begin{equation}
B(p) = \frac{1}{N\, Q} \sum_{q=1}^Q Z_q(p) \;,
\end{equation}
where $Z_q(p)$ indicates the size of the largest
cluster in the network observed, during the $q$th
realization of the Monte Carlo algorithm, when the
bond occupation probability equals $p$. The susceptibility
$\chi_\textrm{B}$ is instead evaluated as
\begin{equation}
\chi_\textrm{B}(p) = N\; \frac{1/(N^2\, Q)\, \sum_{q=1}^Q Z_q(p)Z_q(p) - [B(p)]^2}{B(p)} \;.
\end{equation}
The numerical value of $p_\textrm{c}(N)$ is given by the value of
$p$ for which $\chi_\textrm{B}$ is maximum.
In our simulations,
we also keep track of the size $z$ of all other
clusters present in the network, and monitor the
average size of finite clusters $\langle b \rangle =
\sum_i z_i^2 / \sum_i z_i$, where the sum runs over all
clusters excluding the largest one.
Results shown in the paper are obtained by
considering $Q=10,000$ in simulations
of the percolation process in real networks (Fig.~1),
and $Q=1,000$ (Figs.~2 and~3)
or $Q=100$ (Fig.~4) in artificial graphs.

Simulations for the site percolation model are
performed in a similar way as described above
for the bond percolation model.
The initial configuration is given by a network
with no  nodes, i.e., $n=0$.
Vertices are then sequentially introduced in the network
in a random order. The occupation probability
is defined as $p = n/N$, with $n$ number of
nodes added in the Monte Carlo algorithm. The
definitions
of the order parameter, the susceptibility
and the average cluster size are identical
to those of the bond percolation model.
These quantities are respectively denoted as $S$, $\chi_\textrm{S}$
and $\langle s \rangle$.

\subsection{Random networks}
The generation of a single instance
of the Erd\H{o}s-R\'enyi model with $N$ nodes
and average degree $\langle k \rangle$
is obtained by
connecting each pair of nodes with probability
$\langle k \rangle / (N-1)$.

To generate a random network with $N$ nodes
and power-law degree distribution
\begin{equation}
P(k) \left\{
\begin{array}{ll}
  \sim k^{-\gamma} & \textrm{ , if } k\in[3,\sqrt{N}]
  \\
  = 0 & \textrm{ , otherwise }
\end{array} \right. \;,
\label{eq:degree}
\end{equation}
we make use of the
so-called uncorrelated configuration
model~\cite{molloy1995critical, catanzaro2005generation}.
The support of the degree distribution is chosen
in such a way that the resulting network has no
degree-degree correlations, and is
always composed of a single
connected component. In the generation of
a single instance of
the network model, we first
assign degrees to the nodes according
to the prescribed
$P(k)$. Then, we attach pairs of nodes
at random, preserving their
pre-imposed degrees, but not allowing for
multiple connections and self-loops.

To generate a random network with $N$ nodes
and nonvanishing clustering coefficient,
we make use of the
generalization of the uncorrelated 
configuration model proposed
by Newman~\cite{newman2009random}.
We first assign each node to a number of triangles
randomly extracted from  the 
power-law  distribution
\begin{equation}
P(t) \left\{
\begin{array}{ll}
  \sim t^{-\gamma} & \textrm{ , if } t\in[2,\sqrt{N}]
  \\
  = 0 & \textrm{ , otherwise }
\end{array} \right. \;.
\label{eq:degree2}
\end{equation}
The support of the distribution is chosen
in such a way that the resulting network has no
degree-degree correlations, and is
always composed of a single
connected component. 
After each node has assigned a number $t$, we then
attach triplets of nodes
at random, preserving their
pre-imposed number of triangles $t$, 
but not allowing for
multiple connections and self-loops.
The procedure generates a graph with power-law degree
distribution with degree exponent $\gamma$, and average
clustering coefficient $C \simeq 0.24$ for all sizes $N$.
The clustering coefficient $C$ of the network is
defined as the average value of the clustering
coefficients of all the nodes in the graph.
The clustering coefficient $C_i$ for node $i$ is
defined as 
\begin{equation}
C_ i = \frac{\sum_{r,s} \, A_{i,r} A_{i,s} A_{r,s} } {\sum_{r,s} \, A_{i,r} A_{i,s}} \;,
\end{equation}
with $A_{i,j} = 1$ if nodes $i$ and $j$ are connected, 
and $A_{i,j} =0 $, otherwise.

Results appearing in the paper are obtained on
single network instances. 

\subsection{Finite-size scaling analysis}
On a finite network of size $N$,
the order parameter $B$
follows the scaling
\begin{equation}
  B = N^{-\beta_\textrm{b}/\nu} \, F\left(|p-p_\textrm{c}| \, N^{1/\nu} \right) \;,
  \label{eq:finite}
\end{equation}
where $\beta_\textrm{b}$ is the critical
exponent that regulates the power-law behavior of
$B$ in the infinite-size limit,
$\nu$ is the critical exponent associated with
the correlation length of the system, and $F$ is a
scaling function. The exponent $\nu$ can
be determined by monitoring how the pseudo-critical
threshold $p_\textrm{c}(N)$ changes as a function of the
network size. This quantity is determined by looking
at the location of the peak of the
susceptibility $\chi_\textrm{B}$. 
The pseudo-critical threshold decays
towards the percolation threshold $p_\textrm{c}$
as
\begin{equation}
  p_\textrm{c}(N) -  p_\textrm{c} = a \, N^{-1/\nu} \;. 
  \label{eq:finite3}
  \end{equation}
 If one measures the value of
the order parameter
$B$ at $p = p_\textrm{c}(N)$, the argument of the universal
function does not longer contain any
dependence on either $N$ and $p$, so that
$B \sim N^{-\beta_\textrm{b}/\nu}$, and
the ratio of the critical exponents $\beta_\textrm{b}$ and $\nu$
can be determined from the
decay of the order parameter $B$ for different network sizes.
By definition, the susceptibility $\chi_\textrm{B}$
diverges at pseudo-criticality as $N^{1 - \beta_\textrm{b}/\nu}$. 
The same exact technique can be also used to
determine the power-law scaling of the average cluster
size $\langle b \rangle$. In the case of standard
percolation transitions, the
average cluster size is expected to diverge
at (pseudo-) criticality
as $\langle b \rangle \sim N^{\omega/\nu}$.
Critical exponents for the
site percolation model are numerically
determined in the same way as described above.

\subsection{Percolation thresholds and critical exponents}
For a finite random network
obeying the locally tree-like ansatz,
and with degree distribution $P(k)$,
the pseudo-critical percolation threshold is 
determined as
\begin{equation}
p_\textrm{c}(N) = \frac{\langle k \rangle}{\langle k^2 \rangle
  - \langle k \rangle} \;,
\end{equation}
with $\langle k \rangle = \sum_k k \, P(k)$ and $\langle k^2 \rangle = \sum_k k^2 \, P(k)$ respectively equal
to the first and second moments of the
degree distribution $P(k)$~\cite{callaway2000network, cohen2000resilience}.
This expression is computed with the
so-called heterogeneous mean-field theory. It allows us
to determine the percolation threshold
$p_\textrm{c}$ for networks with
infinite sizes, and also the value of the critical
exponent $\nu$ depending on how $p_\textrm{c}(N)$ approaches $p_\textrm{c}$ as
$N$ grows.
If the degree
distribution is given by Eq.~(\ref{eq:degree}), then
we have $\langle k \rangle = c\, \sum_{k=3}^{\sqrt{N}} k^{1-\gamma}$ and
$\langle k^2 \rangle = c\, \sum_{k=3}^{\sqrt{N}} k^{2-\gamma}$, with
$c$ normalization constant. We have
therefore different predictions based on the value of $\gamma$, i.e.,
depending on whether the second moment of the
distribution is diverging or not as $N$ increases.
For the percolation threshold, we have
\begin{equation}
p_\textrm{c} \left\{
\begin{array}{ll}
  = 0 & \textrm{ , if  } 2 < \gamma \leq 3
  \\
  > 0 & \textrm{ , if  } \gamma > 3
\end{array}
\right. \; .
\end{equation}
For the critical exponent $\nu$, we instead have
\begin{equation}
\nu = \left\{
\begin{array}{ll}
  2/(3 - \gamma)  & \textrm{ , if  } 2 < \gamma < 3
  \\
  (\gamma-1)/(\gamma-3)  & \textrm{ , if  } 3 < \gamma \leq 4
  \\
  3 & \textrm{ , if  } \gamma \geq 4
\end{array}
\right. \; .
\end{equation}
$\gamma=3$ is a pathological case where
we do not expect a power-law decay of $p_\textrm{c}(N)$ to $p_\textrm{c}$, but
rather an exponential one. The prediction in the regime
$2 < \gamma < 3$ is obtained by
accounting for the divergence of the second
moment of the degree distribution with
cutoff given by $\sqrt{N}$.
The prediction in the regime $3 < \gamma \leq 4$
has been obtained by
Wu {\it et al.}~\cite{wu2007numerical}.
For $\gamma \geq 4$ instead,
the exponent $\nu$ equals its mean-field value. 

The estimates of the critical exponent $\beta$ for
the percolation strength are instead given by
\begin{equation}
\beta = \left\{
\begin{array}{ll}
  1/(3 - \gamma) & \textrm{ , if  } 2 < \gamma < 3
  \\
  1 /(\gamma-3)  & \textrm{ , if  } 3 < \gamma \leq 4
  \\
  1 & \textrm{ , if  } \gamma \geq 4
\end{array}
\right. \; .
\end{equation}
These predictions have been obtained by
Cohen {\it et al.}~\cite{cohen2002percolation}.
In the regime $\gamma \geq 4$, $\beta$ assumes its mean-field
value. The results of our
simulations show the prediction in the
regime $2 < \gamma < 3$ to be valid only for the bond percolation
model, i.e., $\beta_\textrm{b} = 1/(3 - \gamma)$.
For the site percolation model, we have instead
$\beta_\textrm{s} = \beta_\textrm{b} + 1 = (4-\gamma)/(3-\gamma)$.

According to our arguments, the exponents $\tau$ and $\omega$,
respectively used to characterize the distribution of cluster
sizes and the average cluster size, are not defined in the
regime $2 < \gamma < 3$, where these quantities
do not obey power-law scalings. They are instead
well defined for $\gamma >3$, where Cohen
{\it et al.}~\cite{cohen2002percolation} predicted
\begin{equation}
\tau = \left\{
\begin{array}{ll}
  (2\gamma - 3)/(\gamma-2) & \textrm{ , if  } 3 < \gamma \leq 4
  \\
  5/2 & \textrm{ , if  } \gamma \geq 4
\end{array}
\right. \; ,
\end{equation}
and
\begin{equation}
\omega = 1  \quad \textrm{ , if  } \gamma > 3
\; .
\end{equation}
Again, the values of the critical exponents for $\gamma>4$ are given
by their mean-field expectations. 
We stress also that the critical exponents are related
by precise hyperscaling relationships. For example, we must
have $2\beta/\nu + \omega/\nu = 1$.

%\bibliography{biblio}

\begin{thebibliography}{10}
\expandafter\ifx\csname url\endcsname\relax
  \def\url#1{\texttt{#1}}\fi
\expandafter\ifx\csname urlprefix\endcsname\relax\def\urlprefix{URL }\fi
\providecommand{\bibinfo}[2]{#2}
\providecommand{\eprint}[2][]{\url{#2}}

\bibitem{stauffer1991introduction}
\bibinfo{author}{Stauffer, D.} \& \bibinfo{author}{Aharony, A.}
\newblock \emph{\bibinfo{title}{Introduction {T}o {P}ercolation {T}heory}}
  (\bibinfo{publisher}{Taylor and Francis}, \bibinfo{year}{1991}).

\bibitem{bollobas2006percolation}
\bibinfo{author}{Bollobas, B.} \& \bibinfo{author}{Riordan, O.}
\newblock \emph{\bibinfo{title}{Percolation}} (\bibinfo{publisher}{Cambridge
  University Press}, \bibinfo{year}{2006}).

\bibitem{callaway2000network}
\bibinfo{author}{Callaway, D.~S.}, \bibinfo{author}{Newman, M.~E.},
  \bibinfo{author}{Strogatz, S.~H.} \& \bibinfo{author}{Watts, D.~J.}
\newblock \bibinfo{title}{Network robustness and fragility: percolation on
  random graphs}.
\newblock \emph{\bibinfo{journal}{Phys. {R}ev. {L}ett.}}
  \textbf{\bibinfo{volume}{85}}, \bibinfo{pages}{5468} (\bibinfo{year}{2000}).

\bibitem{moore2000exact}
\bibinfo{author}{Moore, C.} \& \bibinfo{author}{Newman, M. E.~J.}
\newblock \bibinfo{title}{Exact solution of site and bond percolation on
  small-world networks}.
\newblock \emph{\bibinfo{journal}{Phys. Rev. E}} \textbf{\bibinfo{volume}{62}},
  \bibinfo{pages}{7059--7064} (\bibinfo{year}{2000}).

\bibitem{cohen2002percolation}
\bibinfo{author}{Cohen, R.}, \bibinfo{author}{Ben-Avraham, D.} \&
  \bibinfo{author}{Havlin, S.}
\newblock \bibinfo{title}{Percolation critical exponents in scale-free
  networks}.
\newblock \emph{\bibinfo{journal}{Phys. Rev. E}} \textbf{\bibinfo{volume}{66}},
  \bibinfo{pages}{036113} (\bibinfo{year}{2002}).

\bibitem{dorogovtsev2008critical}
\bibinfo{author}{Dorogovtsev, S.~N.}, \bibinfo{author}{Goltsev, A.~V.} \&
  \bibinfo{author}{Mendes, J.~F.}
\newblock \bibinfo{title}{Critical phenomena in complex networks}.
\newblock \emph{\bibinfo{journal}{Rev. {M}od. {P}hys.}}
  \textbf{\bibinfo{volume}{80}}, \bibinfo{pages}{1275} (\bibinfo{year}{2008}).

\bibitem{PhysRevLett.113.208702}
\bibinfo{author}{Karrer, B.}, \bibinfo{author}{Newman, M. E.~J.} \&
  \bibinfo{author}{Zdeborov\'a, L.}
\newblock \bibinfo{title}{Percolation on sparse networks}.
\newblock \emph{\bibinfo{journal}{Phys. Rev. Lett.}}
  \textbf{\bibinfo{volume}{113}}, \bibinfo{pages}{208702}
  (\bibinfo{year}{2014}).

\bibitem{radicchi2015percolation}
\bibinfo{author}{Radicchi, F.}
\newblock \bibinfo{title}{Percolation in real interdependent networks}.
\newblock \emph{\bibinfo{journal}{Nature Phys.}} \textbf{\bibinfo{volume}{11}},
  \bibinfo{pages}{597--602} (\bibinfo{year}{2015}).

\bibitem{PhysRevLett.113.208701}
\bibinfo{author}{Hamilton, K.~E.} \& \bibinfo{author}{Pryadko, L.~P.}
\newblock \bibinfo{title}{Tight lower bound for percolation threshold on an
  infinite graph}.
\newblock \emph{\bibinfo{journal}{Phys. Rev. Lett.}}
  \textbf{\bibinfo{volume}{113}}, \bibinfo{pages}{208701}
  (\bibinfo{year}{2014}).

\bibitem{PhysRevE.91.052807}
\bibinfo{author}{Faqeeh, A.}, \bibinfo{author}{Melnik, S.} \&
  \bibinfo{author}{Gleeson, J.~P.}
\newblock \bibinfo{title}{Network cloning unfolds the effect of clustering on
  dynamical processes}.
\newblock \emph{\bibinfo{journal}{Phys. Rev. E}} \textbf{\bibinfo{volume}{91}},
  \bibinfo{pages}{052807} (\bibinfo{year}{2015}).

\bibitem{PhysRevE.91.010801}
\bibinfo{author}{Radicchi, F.}
\newblock \bibinfo{title}{Predicting percolation thresholds in networks}.
\newblock \emph{\bibinfo{journal}{Phys. Rev. E}} \textbf{\bibinfo{volume}{91}},
  \bibinfo{pages}{010801} (\bibinfo{year}{2015}).

\bibitem{albert2000error}
\bibinfo{author}{Albert, R.}, \bibinfo{author}{Jeong, H.} \&
  \bibinfo{author}{Barab{\'a}si, A.-L.}
\newblock \bibinfo{title}{Error and attack tolerance of complex networks}.
\newblock \emph{\bibinfo{journal}{Nature}} \textbf{\bibinfo{volume}{406}},
  \bibinfo{pages}{378--382} (\bibinfo{year}{2000}).

\bibitem{cohen2000resilience}
\bibinfo{author}{Cohen, R.}, \bibinfo{author}{Erez, K.},
  \bibinfo{author}{Ben-Avraham, D.} \& \bibinfo{author}{Havlin, S.}
\newblock \bibinfo{title}{Resilience of the internet to random breakdowns}.
\newblock \emph{\bibinfo{journal}{Phys. Rev. Lett.}}
  \textbf{\bibinfo{volume}{85}}, \bibinfo{pages}{4626} (\bibinfo{year}{2000}).

\bibitem{molloy1995critical}
\bibinfo{author}{Molloy, M.} \& \bibinfo{author}{Reed, B.}
\newblock \bibinfo{title}{A critical point for random graphs with a given
  degree sequence}.
\newblock \emph{\bibinfo{journal}{Random Struct. Algor.}}
  \textbf{\bibinfo{volume}{6}}, \bibinfo{pages}{161--180}
  (\bibinfo{year}{1995}).

\bibitem{catanzaro2005generation}
\bibinfo{author}{Catanzaro, M.}, \bibinfo{author}{Bogu{\~n}{\'a}, M.} \&
  \bibinfo{author}{Pastor-Satorras, R.}
\newblock \bibinfo{title}{Generation of uncorrelated random scale-free
  networks}.
\newblock \emph{\bibinfo{journal}{Phys. Rev. E}} \textbf{\bibinfo{volume}{71}},
  \bibinfo{pages}{027103} (\bibinfo{year}{2005}).

\bibitem{newman2000efficient}
\bibinfo{author}{Newman, M. E.~J.} \& \bibinfo{author}{Ziff, R.}
\newblock \bibinfo{title}{Efficient monte carlo algorithm and high-precision
  results for percolation}.
\newblock \emph{\bibinfo{journal}{Phys. Rev. Lett.}}
  \textbf{\bibinfo{volume}{85}}, \bibinfo{pages}{4104} (\bibinfo{year}{2000}).

\bibitem{wu2007numerical}
\bibinfo{author}{Wu, Z.} \emph{et~al.}
\newblock \bibinfo{title}{Numerical evaluation of the upper critical dimension
  of percolation in scale-free networks}.
\newblock \emph{\bibinfo{journal}{Phys. Rev. E}} \textbf{\bibinfo{volume}{75}},
  \bibinfo{pages}{066110} (\bibinfo{year}{2007}).

\bibitem{dorogovtsev2010lectures}
\bibinfo{author}{Dorogovtsev, S.~N.}
\newblock \emph{\bibinfo{title}{Lectures On Complex Networks}},
  vol.~\bibinfo{volume}{24} (\bibinfo{publisher}{Oxford University Press
  Oxford}, \bibinfo{year}{2010}).

\bibitem{albert2002statistical}
\bibinfo{author}{Albert, R.} \& \bibinfo{author}{Barab{\'a}si, A.-L.}
\newblock \bibinfo{title}{Statistical mechanics of complex networks}.
\newblock \emph{\bibinfo{journal}{Rev. Mod. Phys.}}
  \textbf{\bibinfo{volume}{74}}, \bibinfo{pages}{47} (\bibinfo{year}{2002}).

\bibitem{PhysRevE.76.045101}
\bibinfo{author}{Newman, M. E.~J.}
\newblock \bibinfo{title}{Component sizes in networks with arbitrary degree
  distributions}.
\newblock \emph{\bibinfo{journal}{Phys. Rev. E}} \textbf{\bibinfo{volume}{76}},
  \bibinfo{pages}{045101} (\bibinfo{year}{2007}).

\bibitem{serrano2011percolation}
\bibinfo{author}{Serrano, M.~A.}, \bibinfo{author}{Krioukov, D.} \&
  \bibinfo{author}{Bogu\~n\'a, M.}
\newblock \bibinfo{title}{Percolation in self-similar networks}.
\newblock \emph{\bibinfo{journal}{Phys. Rev. Lett.}}
  \textbf{\bibinfo{volume}{106}}, \bibinfo{pages}{048701}
  (\bibinfo{year}{2011}).

\bibitem{Trugman86}
\bibinfo{author}{Trugman, S.~A.} \& \bibinfo{author}{Weinrib, A.}
\newblock \bibinfo{title}{Percolation with a threshold at zero: a new
  universality class}.
\newblock \emph{\bibinfo{journal}{Phys. Rev. B}} \textbf{\bibinfo{volume}{31}},
  \bibinfo{pages}{2974--2980} (\bibinfo{year}{1985}).

\bibitem{barabasi1999emergence}
\bibinfo{author}{Barab{\'a}si, A.-L.} \& \bibinfo{author}{Albert, R.}
\newblock \bibinfo{title}{Emergence of scaling in random networks}.
\newblock \emph{\bibinfo{journal}{Science}} \textbf{\bibinfo{volume}{286}},
  \bibinfo{pages}{509--512} (\bibinfo{year}{1999}).

\bibitem{pastor2001epidemic}
\bibinfo{author}{Pastor-Satorras, R.} \& \bibinfo{author}{Vespignani, A.}
\newblock \bibinfo{title}{Epidemic spreading in scale-free networks}.
\newblock \emph{\bibinfo{journal}{Phys. Rev. Lett.}}
  \textbf{\bibinfo{volume}{86}}, \bibinfo{pages}{3200} (\bibinfo{year}{2001}).

\bibitem{newman2002spread}
\bibinfo{author}{Newman, M.~E.}
\newblock \bibinfo{title}{Spread of epidemic disease on networks}.
\newblock \emph{\bibinfo{journal}{Phys. Rev. E}} \textbf{\bibinfo{volume}{66}},
  \bibinfo{pages}{016128} (\bibinfo{year}{2002}).

\bibitem{pastor2014epidemic}
\bibinfo{author}{Pastor-Satorras, R.}, \bibinfo{author}{Castellano, C.},
  \bibinfo{author}{Van~Mieghem, P.} \& \bibinfo{author}{Vespignani, A.}
\newblock \bibinfo{title}{Epidemic processes in complex networks}.
\newblock \emph{\bibinfo{journal}{Rev. Mod. Phys.}}
  \textbf{\bibinfo{volume}{87}}, \bibinfo{pages}{925--979}
  (\bibinfo{year}{2015}).

\bibitem{kitano2004biological}
\bibinfo{author}{Kitano, H.}
\newblock \bibinfo{title}{Biological robustness}.
\newblock \emph{\bibinfo{journal}{Nat. Rev. Genet.}}
  \textbf{\bibinfo{volume}{5}}, \bibinfo{pages}{826--837}
  (\bibinfo{year}{2004}).

\bibitem{watts1998collective}
\bibinfo{author}{Watts, D.~J.} \& \bibinfo{author}{Strogatz, S.~H.}
\newblock \bibinfo{title}{Collective dynamics of 'small-world' networks}.
\newblock \emph{\bibinfo{journal}{Nature}} \textbf{\bibinfo{volume}{393}},
  \bibinfo{pages}{440--442} (\bibinfo{year}{1998}).

\bibitem{newman2009random}
\bibinfo{author}{Newman, M.~E.}
\newblock \bibinfo{title}{Random graphs with clustering}.
\newblock \emph{\bibinfo{journal}{Phys. Rev. Lett.}}
  \textbf{\bibinfo{volume}{103}}, \bibinfo{pages}{058701}
  (\bibinfo{year}{2009}).

\bibitem{PhysRevE.66.056105}
\bibinfo{author}{Warren, C.~P.}, \bibinfo{author}{Sander, L.~M.} \&
  \bibinfo{author}{Sokolov, I.~M.}
\newblock \bibinfo{title}{Geography in a scale-free network model}.
\newblock \emph{\bibinfo{journal}{Phys. Rev. E}} \textbf{\bibinfo{volume}{66}},
  \bibinfo{pages}{056105} (\bibinfo{year}{2002}).

\bibitem{PhysRevE.81.066114}
\bibinfo{author}{Gleeson, J.~P.}, \bibinfo{author}{Melnik, S.} \&
  \bibinfo{author}{Hackett, A.}
\newblock \bibinfo{title}{How clustering affects the bond percolation threshold
  in complex networks}.
\newblock \emph{\bibinfo{journal}{Phys. Rev. E}} \textbf{\bibinfo{volume}{81}},
  \bibinfo{pages}{066114} (\bibinfo{year}{2010}).

\bibitem{ripeanu2002mapping}
\bibinfo{author}{Ripeanu, M.}, \bibinfo{author}{Foster, I.} \&
  \bibinfo{author}{Iamnitchi, A.}
\newblock \bibinfo{title}{Mapping the gnutella network: properties of
  large-scale peer-to-peer systems and implications for system design}.
\newblock \emph{\bibinfo{journal}{IEEE Internet Comput.}}
  \textbf{\bibinfo{volume}{6}}, \bibinfo{pages}{2002} (\bibinfo{year}{2002}).

\bibitem{leskovec2007graph}
\bibinfo{author}{Leskovec, J.}, \bibinfo{author}{Kleinberg, J.} \&
  \bibinfo{author}{Faloutsos, C.}
\newblock \bibinfo{title}{Graph evolution: densification and shrinking
  diameters}.
\newblock In \emph{\bibinfo{booktitle}{ACM Transactions on Knowledge Discovery
  from Data (TKDD)}}, vol.~\bibinfo{volume}{1}, \bibinfo{pages}{2}
  (\bibinfo{publisher}{ACM}, \bibinfo{year}{2007}).

\bibitem{leskovec2005graphs}
\bibinfo{author}{Leskovec, J.}, \bibinfo{author}{Kleinberg, J.} \&
  \bibinfo{author}{Faloutsos, C.}
\newblock \bibinfo{title}{Graphs over time: densification laws, shrinking
  diameters and possible explanations}.
\newblock In \emph{\bibinfo{booktitle}{KDD '05 Proceedings of the eleventh ACM
  SIGKDD international conference on Knowledge discovery in data mining}},
  \bibinfo{pages}{177--187} (\bibinfo{organization}{ACM},
  \bibinfo{year}{2005}).

\end{thebibliography}

\end{document}